\documentstyle[prl,aps,preprint]{revtex}
\begin{document}

\title{ Two-hole bound states in modified $t$-$J$ model}

\author {M.Yu. Kuchiev\cite{Ioffe} and O. P. Sushkov\cite{Budker}}
\address { School of Physics, The University of New South Wales
 Sydney 2052, Australia}

\maketitle

\begin{abstract}
We consider modified $t-J$ model with minimum of single-hole
dispersion at the points $(0,\pm \pi)$, $(\pm \pi,0)$.
It is shown that two holes on antiferromagnetic background
produce a bound state which properties strongly differs
from the states known in the unmodified $t-J$ model. The bound
state is d-wave, it  has four nodes on the face
of the magnetic Brillouin zone. However, in the coordinate representation
it looks like as usual s-wave.
\end{abstract}

\vspace{0.5cm}
\hspace{3.cm}PACS: 74.20.-z, 75.30.Ds, 75.50.Ee
\vspace{1.cm}

It is widely believed that essential physics of high-$T_c$
Cooper Oxide superconductors is described by $t-J$ model on two
dimensional square lattice. It is well known that at half-filling
(one hole per site) this model is equivalent to
the Heisenberg antiferromagnetic insulator with the long-range
antiferromagnetic order in the ground state.
However any finite doping by holes destroys the long-range
order (see e.g. Refs. \cite{Shr9,Sin0,Ede1,Iga2,Sus3}).
The problem of two-hole bound state is closely related to the problem of
superconducting pairing. However it is substantially simpler
because no destruction of the
antiferromagnetic order is to be considered,
the two holes can not destroy the
well defined antiferromagnetic background.

Single hole (single hole above half-filling) properties in the
$t$-$J$ model are by now well
established. A hole is an object with complex structure due to
virtual admixture of spin excitations. The bottom of the hole
band is at the momenta ${\bf k}=(\pm \pi/2, \pm \pi/2)$. The dispersion
is almost degenerate along the face of the magnetic Brillouin zone
$\gamma_{\bf k}={1 \over 2}(\cos k_x+\cos k_y)=0$  (see e.g.
Refs.\cite{Tru8,Sch8,Shr8,Kan9,Bon9,Dag0,Mart1,Liu2,Sus2,Gia3}).
So the dispersion at the minima ${\bf k}=(\pm \pi/2, \pm \pi/2)$
is very anisotropic: the effective  mass is very large
along the face of magnetic Brillouin zone, in contrast
it is equal to approximately  two
electron masses in the perpendicular direction. The mass ratio is
about 6\cite{Mart1,Gia3}. A hole can be characterized by the
z-projection of spin $S_z= \pm {1\over2}$, where z is a direction of the
staggered magnetization.

In the $t-J$ model two holes with opposite $S_z$ produce an
infinite set of bound states. Classification of these states as well
as calculation of their binding energies is given in our
work\cite{Kuch3}. The interesting peculiarity of these states
is an unusual relation between the wave function in ${\bf k}$-representation,
parity, and the wave function in coordinate representation. For
example the state which is usually called d-wave possesses four nodes on the
face of magnetic Brillouin zone and  has positive parity, but
in the coordinate representation it looks like a p-wave.
This peculiarity is due to the existence of the antiferromagnetic
background. For $t/J \le 2-3$
there are also the short-range bound states
whose size is close to a lattice spacing. They belong to d- and p-wave
symmetry according to
conventional classification by the number of
nodes on the face of magnetic Brillouin zone. They are known as well
from computer
simulations \cite{Ede2,Bon2,Poilblanc,Sher,PoilDag} and from
variational calculations  \cite{Cher3}.
Usually it is convenient to consider these latter states
in the momentum representation. That is why the unusual
properties of the wave function in the momentum and coordinate
representations are not so interesting for these deep states.

 The pure $t-J$ model gives a Fermi surface with small hole pockets
centred at $(\pm \pi/2, \pm \pi/2)$. The mentioned above properties of
two-hole bound states are essentially based on the fact that
minima of single-hole dispersion are at these points.
However recent experimental data on angle-resolved photoemission
(see e.g. Refs.\cite{Des3,Des33}) demonstrate
the existence of large Fermi surface
with minima at the points $(\pm \pi,0)$, $(0,\pm \pi)$.
Let us incorporate these data into the $t-J$ model.
With this purpose
let us include into the Hamiltonian a small hopping
amplitude along the plaquette diagonals (compare with Ref.\cite{Dag4}).
So let us consider
$t-t^{\prime}-J$ model. We will consider very small values of diagonal
hopping ($t^{\prime} \le 0.1t$), therefore all  properties of hole-hole and
hole spin-wave interactions remain practically unchanged in comparison with
pure $t-J$ model. However this small $t^{\prime}$ is enough to shift
minima of single-hole dispersion from $(\pm \pi/2, \pm \pi/2)$ to
$(\pm \pi,0)$, $(0,\pm \pi)$, and we will see that it drastically
changes the properties of two-hole bound states.
In the modified model
there is only   one exponentially shallow bound
state of d-wave symmetry
having four nodes on the face of the Brillouin zone.
In the coordinate representation
the wave-function of this state possesses no nodes which
makes it look like  s-wave.

 In $t-J$ model for $t \le 5$ the hole dispersion
can be well approximated by the following analytical
expression\cite{Sus2,Suhf} (we set below $J=1$)
\begin{equation}
\label{disp}
\epsilon_{\bf k}=-\sqrt{\Delta^2/4+4t^2(1+y)-4t^2(x+y)\gamma_{\bf k}^2}
+{1\over4}\beta_2(\cos k_x -\cos k_y)^2,
\end{equation}
where $ \Delta \approx 1.33$, $x \approx 0.56$, $y \approx 0.14$.
The parameters $\Delta,x,y$ are some combinations of the ground
state spin correlators\cite{Sus2}. The $\beta_2$ term gives the
dispersion along the face of the Brillouin zone. It is rather small and
according to Refs.\cite{Mart1,Gia3} can be estimated as
$\beta_2 \sim 0.1 t$ at $\Delta/4 \le t \le 5$. The minima of
the dispersion (\ref{disp}) are at the points $(\pm \pi/2, \pm \pi/2)$,
and the mass ratio for different directions at these points is about 6.

In $t-t^{\prime}-J$ model we have to add to the dispersion (\ref{disp})
the diagonal hopping term $4t^{\prime}_{eff}\cos k_x \cos k_y$. The
effective hopping parameter $t^{\prime}_{eff}$ differs from the bare one
due to the dressing of the hole by spin-wave excitations. For
$t^{\prime} \ll t$ and $\Delta/4 \le t \le 5$ one gets the estimation
$t^{\prime}_{eff} \sim 0.5t^{\prime}$. The correction to dispersion
can be represented as $4t^{\prime}_{eff}\cos k_x \cos k_y
= 4t^{\prime}_{eff}[\gamma_{\bf k}^2-{1\over 4}(\cos k_x - \cos k_y)^2]$.
The $\gamma_{\bf k}^2$ term can be neglected in comparison with
that in Eq.(\ref{disp}), and the second term gives the renormalization
of dispersion along the face of zone: $\beta_2 \to \beta_2-4t^{\prime}_{eff}$.
So the only difference of the modified $t-J$ model from the
pure one is a different
value of $\beta_2$. Therefore we can
consider $\beta_2$ as independent parameter
instead of $t^{\prime}$. For positive $\beta_2$
the dispersion minima are at $(\pm \pi/2, \pm \pi/2)$, while for
negative $\beta_2$ the dispersion minima are at the points
$(\pm \pi,0)$,$(0,\pm \pi)$.

The state of two holes with zero total momentum and zero
projection of spin is of the form:
$|2h \rangle =  \sum_{\bf k} g_{\bf k}
 h^{\dag}_{{\bf k}\uparrow} h^{\dag}_{- {\bf k} \downarrow} |0\rangle$.
Here $|0\rangle$ is a wave function of the quantum antiferromagnet
(ground state of Heisenberg model), $h^{\dag}_{{\bf k}\sigma}$ is a
creation operator of a dressed hole with dispersion (\ref{disp}) and
$\sigma= \pm {1\over 2}$.
To avoid misunderstanding note that a number of magnons in
a dressed hole is not limited. Very roughly at $t/J=3$ the
weight of the bare hole in $h^{\dag}_{{\bf k}\sigma}$ is about $25\%$,
the weight of configurations ``bare hole + 1 magnon'' is $\sim 50\%$,
and of configurations ``bare hole + 2 or more magnons'' is $\sim 25\%$.
The wave-function $g_{{\bf k}}$ satisfies  the Bethe-Salpeter equation
which is convenient to write for $\chi_{\bf k}=
(E-2\epsilon_{\bf k}) g_{\bf k}$.
\begin{equation}
\label{BS}
 \chi_{\bf k} = \sum_{\bf k^{\prime}}
{{\Gamma_{{\bf k}{\bf k^{\prime}}}
\chi_{\bf k^{\prime}}}\over{E-2\epsilon_{\bf k^{\prime}}}}.
\end{equation}
Summation over ${\bf k^{\prime}}$ is restricted inside the magnetic
Brillouin zone ($\gamma_{\bf k^{\prime}} \ge 0$).
The kernel $\Gamma_{{\bf k}{\bf k^{\prime}}}$ was derived in
the work\cite{Kuch3}. It is the sum of the
single spin-wave exchange contribution and the ``contact'' part:
$\Gamma=\Gamma_{sw}+\Gamma_{contact}$.
The single spin-wave exchange part is of the form
\begin{equation}
\label{Gamsw}
\Gamma_{sw}=
-16f^2 {{(\gamma_{\bf -k}u_{\bf q}+\gamma_{\bf k^{\prime}}v_{\bf q})
(\gamma_{\bf -k^{\prime}}u_{\bf q}+\gamma_{\bf k}v_{\bf q})}
\over{E-\epsilon_{\bf k^{\prime}}-\epsilon_{\bf k}-\omega_{\bf q}}},
\end{equation}
where ${\bf q}={\bf k^{\prime}}+{\bf k}$ is the momentum of virtual spin-wave,
$\omega_{\bf q}=2\sqrt{1-\gamma_{\bf q}^2}$ is the frequency of this
wave, $u_{\bf q}=\sqrt{{1\over{\omega_{\bf q}}}+{1\over 2}}$ and
$v_{\bf q}=-sign(\gamma_{\bf q})\sqrt{{1\over{\omega_{\bf q}}}-{1\over 2}}$
are the parameters of the Bogoliubov transformation diagonalizing spin-wave
Hamiltonian. Hole spin-wave coupling constant $f$ is a function of $t$.
It was calculated in the work\cite{Suhf}.
The effective interaction (\ref{Gamsw}) is given by the second order
of perturbation theory in
quasihole-spin-wave interaction. Let us stress that even for $t > J$ the
quasihole-spin-wave interaction has the same form as
for $t \ll J$ ({\em i.e.} as for bare hole  operators)
with an added renormalization factor which is of the order $J/t$ for
$t \gg J$.
This remarkable property of the $t$-$J$ model is due to the absence
of the single-loop correction to the hole-spin-wave vertex. It
was first stated implicitly in the
work\cite{Kan9}. In Refs.\cite{Mart1,Liu2,Suhf} it was
explicitly demonstrated that the vertex corrections with different kinematic
structure are of the order of few percent at $t/J \approx 3$.
 There is also a weak $q$-dependence of the coupling constant $f$.
The value of $f$ which we use in the present work was calculated
in the Ref.\cite{Suhf} in the long wave length limit: $q=0$.
At $q\sim\pi$, the factor $f$ is 17--20 \% bigger
  than at $q=0$, see the discussion between Eqs.\ (13) and (14) of
Ref.\cite{Kuch3}. This ${\bf q}$-dependence does not influence the
qualitative results of
the present work. However it could be important for some quantitative
results, see conclusion.

 For the states under
consideration the energy $E$ in the denominator (\ref{Gamsw})
is negligible because
$|E| \ll \epsilon_{\bf k^{\prime}}+\epsilon_{\bf k}+\omega_{\bf q}$.
So we can set in (\ref{Gamsw}) $E=0$. Then  Eq.(\ref{BS})
is reduced to the Schroedinger equation.
The contact hole-hole interaction is of the form\cite{Cher3,Kuch3}:
$\Gamma_{contact}\approx
8A\gamma_{{\bf k}-{\bf k^{\prime}}}$.
 The vertices $\Gamma_{sw}$ and $\Gamma_{contact}$
have been derived for pure $t-J$ model. Here we consider $t-t^{\prime}-J$
model. However $t^{\prime} \sim 0.1t \ll t$ and therefore the corrections
are small. Actually they are very small
$\sim (t^{\prime}/t)^2$ because the interactions are due to the spin-flip
processes but diagonal hopping does not result in the spin-flip. So the
interactions $\Gamma_{sw}$ and $\Gamma_{contact}$ derived for the pure $t-J$
model are valid for modified model as well.

For small $t$ the parameter $A$ in $\Gamma_{contact}$
approaches the value  -0.25, which gives the well known hole-hole
attraction induced by the reduction of the number of missing
antiferromagnetic links in the problem. However for realistic
superconductors $t\approx 3$ (see e.g.Refs.\cite{Esk0,BCh3}).
In this region of $t$ the parameter $A$ vanishes due to the hopping
contribution\cite{Cher3}. So this mechanism of attraction is switched off.
In contrast the spin-wave exchange mechanism (\ref{Gamsw}) is negligible
for small $t$ where $f \propto t$, and it is the most important at
large $t$ where $f$ approaches a $t$ independent
constant. We are interested mainly
in ``physical'' values of $t$: $t \approx 3$. Therefore in the present
work we neglect contact interaction ($A=0$) and consider only
the spin-wave exchange (\ref{Gamsw}).

  Let us concentrate here on the lowest waves of positive parity:
d-wave and g-wave. The symmetries of corresponding wave functions
are shown in Fig.1. It was demonstrated in the Ref.\cite{Kuch3}
that in the pure $t-J$ model (positive $\beta_2$) there is an infinite set of
bound states in each of these waves,
and binding energies of the states increase
with decrease of $\beta_2$. Therefore  consider here first the
case $\beta_2=0$. We found no  analytical solution for
this case but numerical integration of the Eq.(\ref{BS}) is straightforward.
The binding energies of four lowest bound states of d- and g-wave symmetry
at $t=3$, $f=1.79$, $\beta_2=0$ are presented in Table 1.
In the same Table we present the values of the radii of
bound states in coordinate representation
$r_{rms}=\sqrt{\langle r^2 \rangle}$, where
$\langle r^2 \rangle = \int r^2 \psi^2({\bf r}) d^2r=
\sum_{\bf k} |{{\partial g_{\bf k}} \over {\partial {\bf k}}}|^2$.
{}From Table 1 we see that at $\beta_2=0$ the short-range d-wave
bound state ($r=1.67$) exists at $t=3$. Recall that in the pure $t-J$
model it vanishes at
$t \approx 2-3$\cite{Ede2,Bon2,Poilblanc,Sher,PoilDag,Cher3}.

The negative value of $\beta_2$ changes the situation drastically.
At $\beta_2 \le \beta_{g}^{crit}$ the
 g-wave bound state disappears. The critical value is
very small. For $t=3$, $f=1.79$ it is $\beta_{g}^{crit}=-0.02$.
In the d-wave at $\beta_2 \le \beta_{d}^{crit}$ only one shallow bound
state survives.
For $t=3$, $f=1.79$ the d-wave critical value
is $\beta_{d}^{crit}=-0.013$.
Let us elucidate the structure of this shallow d-wave state.
Bound state wave function is concentrated around the minima of
hole-dispersion $(0,\pm \pi)$, $(\pm \pi,0)$. Therefore it is
convenient to shift the center of the Brillouin zone to one of these points.
We transfer the usual Brillouin zone centered at ${\bf k} = (0,0)$
(solid square at Fig.2) to the zone centred at ${\bf k_c}=(0,\pi)$
(dashed square at Fig.2). This can be done by a combination of
translations by the vectors of inverse magnetic lattice
${\bf g}=(\pm \pi, \pm \pi)$ which are shown in Fig.2.
The interaction $\Gamma$ changes the sign under the translation
by ${\bf g}$: $\Gamma_{{\bf k}+{\bf g},{\bf k^{\prime}}}
=\Gamma_{{\bf k},{\bf k^{\prime}+{\bf g}}}=-\Gamma_{{\bf k}{\bf k^{\prime}}}$.
This is a consequence of an existence of two magnetic
sublattices. One can verify the change of sign
using explicit formula
(\ref{Gamsw}) for $\Gamma_{sw}$. The wave function obeys similar
relation $g_{{\bf k}+{\bf g}}=-g_{\bf k}$, because it is a solution
of Eq.(\ref{BS}).
 Due to this property starting from a d-wave function
in the usual Brillouin zone (four nodes on the face of zone)
after translation (Fig.2) we get constant sign wave function inside the
shifted zone. The wave function vanishes at the face of this zone.

For convenience let us define ${\bf p}$ be a deviation of a momentum
from the center of shifted zone: ${\bf p}={\bf k}-{\bf k_c}$, and
$\Gamma_{{\bf p}{\bf p^{\prime}}}^c=
\Gamma_{{\bf p}+{\bf k_c},{\bf p^{\prime}}+{\bf k_c}}$.
In the modified model the region of
long distances corresponds to small momenta
${\bf p}$: $r \sim p^{-1} \gg 1$. Using this
fact one can easily prove that the Fourier transformation of
$\Gamma_{{\bf p}{\bf p^{\prime}}}^c$ gives in coordinate space the
short-range potential. This situation is  quite different from the pure $t-J$
model where the region of
long distances corresponds to small deviation
of ${\bf k}$ from $(\pm \pi/2, \pm \pi/2)$ and therefore the same
interaction gives the long-range potential \cite{Kuch3}.

We consider shallow bound state. Therefore let us solve Eq.(\ref{BS})
using perturbation theory. In zero approximation $\chi_{\bf p}^{(0)}=1$.
The integral in the right-hand side of equation (\ref{BS}) is logarithmically
divergent at ${\bf p^{\prime}}=0$ because near this point
$\epsilon_{\bf p^{\prime}} \approx const+{1\over 2}|\beta_2|{\bf p}
^{\prime 2}$. Calculating this integral with logarithmic
accuracy and taking into account
a consistency condition at ${\bf p}=0$ one can
get the well known formula\cite{Land} for exponentially shallow s-wave bound
state in two dimensions :
$E \sim -|\beta_2| \cdot p_{max}^2 \cdot
\exp \bigl(-4\pi \beta_2/\Gamma_{00}^c\bigr)$.
However in our case $\Gamma_{00}^c=0$ and therefore this approach should
be improved.  In order to do this  let us first evaluate
from (\ref{BS}) a correction
containing the big logarithm:
$\chi_{\bf p}^{(1)}=-\Gamma_{{\bf p},0}^c \cdot L$, where
\begin{equation}
\label{ln}
L=\int {1\over{2\epsilon_{\bf p}-E}}{{d^2{\bf p}}\over{(2\pi)^2}}=
{1\over{4\pi |\beta_2|}}\ln \biggl({{\beta_2 p_{max}^2}\over{E}}\biggr),
\end{equation}
$p_{max} \sim 1$. Note that $\chi_{{\bf p}=0}^{(1)}=0$.
Equation (\ref{BS}) gives
\begin{equation}
\label{PT}
1+\chi_{\bf p}^{(1)}+ \chi_{\bf p}^{(2)} = \sum_{\bf p^{\prime}}
{{\Gamma_{{\bf p}{\bf p^{\prime}}}^c
(1+\chi_{\bf p^{\prime}}^{(1)})}
\over{E-2\epsilon_{\bf p^{\prime}}}}.
\end{equation}
Condition of consistency at small ${\bf p}$ (i.e. at large distances)
$\chi_{{\bf p}=0}^{(2)}=0$ together with
Eq.(\ref{PT}) gives the equation for the binding energy
\begin{eqnarray}
\label{QR}
1&=&Q+R \cdot L, \\
Q&=&-\sum_{\bf p}{{\Gamma_{0,{\bf p}}^c}
\over{2\epsilon_{\bf p}}}, \nonumber \\
R&=&\sum_{\bf p}{{\Gamma_{0,{\bf p}}^c\Gamma_{{\bf p},0}^c}
\over{2\epsilon_{\bf p}}}. \nonumber
\end{eqnarray}
Summation over ${\bf p}$ is restricted inside the dashed square in Fig.2.
{}From Eq.(\ref{QR}) one finds
\begin{equation}
\label{E}
E=-|\beta_2| \cdot p_{max}^2 \cdot \exp\biggl(-{{4\pi |\beta_2| (1-Q)}\over R}
\biggr).
\end{equation}
So we get exponentially shallow bound state. The integrals $Q$ and $R$
can be easily calculated. They are $\beta_2$-dependent. For
$t=3$, $f=1.79$ and $-0.36 \le \beta_2 \le 0$ their typical values
are: $0.3 \le Q \le 0.7$, $0.2 \le R \le 0.4$.
The values of exponent in (\ref{E}) for different
$\beta_2$ and $t=3$, $f=1.79$ are presented in Table 2.
In the same Table we present $\ln(\beta_2/E)$ with $E$ obtained by exact
numerical integration of Eq.(\ref{BS}).
The agreement of analytical formula (\ref{E}) with the
exact result is perfect. From comparison we find $p_{max} \approx 1.2$.

We have found exponentially shallow two-hole bound state
in modified $t-J$ model having minima of the single-hole dispersion
at the points $(0,\pm \pi)$, $(\pm \pi,0)$.
There is no such state in the pure $t-J$ model.
Thus state behaves as s-wave in coordinate space
and also as s-wave in momentum space  if the rotations
around the point ${\bf k_c}=(0,\pi)$ are considered. According to
the standard classification this state possesses the symmetry of
d-wave, - it has
four nodes on the face of the usual
magnetic Brillouin zone centered at ${\bf k_c}=(0,0)$.
The size of this state is large
$r_{rms}=\sqrt{\langle r^2 \rangle}=\sqrt{{2\over 3}|\beta_2/E|}
\gg 1$ justifying the analytical formulas used to describe it,
but numerically formula (\ref{E}) remains valid even
if $r_{rms}\approx 2$ when  the  binding energy is not small.

The factor $R$ in the exponent of (\ref{E})
quadratically depends on the interaction potential $\Gamma$.
This causes extremely strong dependence of the binding energy
on the hole spin-wave coupling constant $f$: $E \sim \-exp{(-const/f^4)}$.
For numerical estimations we have used ``infrared'' value of the
coupling constant $f$. However, it is evident  from Eq.(\ref{QR})
that the main contribution comes from large ${\bf p}$, where $f$ is by
17--20\% bigger. So numerically we have substantially underestimated the
binding energy, and account of ${\bf q}$ dependence of running
coupling constant $f$ requires further calculations.

The implication of the found bound state for superconducting pairing will be
considered elsewhere. However we have to note here that starting
from the spin-wave exchange in modified $t-J$ model we get the effective
short-range attraction. This supports the scenario suggested
in the recent work\cite{Dag4} where a short range attraction was introduced
by hands.

{\bf ACKNOWLEDGMENTS}

  We are very grateful to G.A.Sawatzky, V.V.Flambaum, and
D.I.Khomskii for stimulating discussions.
Part of this work has been
done during Workshop on High-$T_c$ Superconductivity
organized by
Australian National Centre for Theoretical Physics at
Australian National University.
We are very grateful to the organizers of Workshop.

\newpage

\tighten

\begin{table}
\caption{Binding energies (in units of $J$) and radii of the four lowest
bound states in d- and g-wave at  $\beta_2=0$, $t=3$, $f=1.79$.}
\label{tab1}
\begin{tabular}{c|ccccc}
        & n    & 1    &  2      & 3     &  4      \\
\hline
d-wave & -E        & 0.106 & $0.17\cdot 10^{-1}$ & $0.70\cdot 10^{-2}$ &
$0.39\cdot 10^{-2}$\\
       & $r_{rms}$ & 1.67  & 5.2  & 7.8 & 10 \\
\hline
g-wave & -E        & $0.33\cdot 10^{-1}$ & $0.10\cdot 10^{-1}$ &
$0.51\cdot 10^{-2}$ & $0.31\cdot 10^{-2}$\\
       & $r_{rms}$ & 4.1  & 6.8  & 9.3 & 12 \\
\end{tabular}
\end{table}

\begin{table}
\caption{Comparison of numerical and analytical calculations
for different $\beta_2$ ($t=3$, $f=1.79$):
The value of $\ln(\beta_2/E)$ with $E$ obtained by numerical
integration of the Schroedinger equation,
and the value of exponent in equation (7).}
\label{tab2}
\begin{tabular}{c|cccccc}
$\beta_2$          & -0.1   & -0.2  &  -0.3 \\
\hline
$\ln(\beta_2/E)$ & 1.0    & 4.5   &  10.3 \\
${{4\pi |\beta_2| (1-Q)}\over{R}}$
                   & 1.3    & 4.3   & 9.9   \\
\end{tabular}
\end{table}

{\bf FIGURE CAPTIONS}

FIG. 1. Symmetry of the bound state wave function for (a) d-wave and
(b) g-wave.\\

FIG. 2. Transition from the usual Brillouin zone $\gamma_{\bf k} \ge 0$
(solid square) to the zone centred at ${\bf k_c}=(0,\pi)$
(dashed square).

\begin{references}
\bibitem[a]{Ioffe}Also at the
A.F.Ioffe Physical-Technical Institute, 194021 St. Petersburg, Russia
\bibitem[b]{Budker} Also at the Budker Institute of Nuclear Physics,
630090 Novosibirsk, Russia

\bibitem{Shr9} B.Shraiman and E.Siggia, Phys. Rev. Lett. {\bf 62}, 1564 (1989).
\bibitem{Sin0} A.Singh and Z.Tesanovic, Phys. Rev. B {\bf 41}, 614 (1990).
\bibitem{Ede1} R.Eder, Phys. Rev. B {\bf 43}, 10706 (1991).
\bibitem{Iga2} J.Igarashi and P.Fulde, Phys. Rev. B {\bf 45}, 10419 (1992).
\bibitem{Sus3} O.P.Sushkov and V.V.Flambaum, Physica C {\bf 206}, 269 (1993).

\bibitem{Tru8} S. A. Trugman, Phys. Rev. B {\bf 37}, 1597 (1988).
\bibitem{Sch8} S. Schmitt-Rink, C. M. Varma, and A. E. Ruckenstein,
 Phys. Rev. Lett. {\bf 60}, 2793 (1988).
\bibitem{Shr8} B. Shraiman and E. Siggia,
 Phys. Rev. Lett. {\bf 61}, 467 (1988).
\bibitem{Kan9} C. L. Kane, P. A. Lee, and N. Read, Phys. Rev. B {\bf 39},
6880 (1989).
\bibitem{Bon9} J. Bonca, P. Prelovsek, and I. Sega, Phys. Rev. B {\bf 39},
7074 (1989).
\bibitem{Dag0} E. Dagotto, R. Joynt, S. Bacci, and E. Cagliano, Phys. Rev. B
  {\bf 41}, 9049 (1990).
\bibitem{Mart1} G. Martinez and P. Horsch, Phys. Rev. B {\bf 44}, 317 (1991).
\bibitem{Liu2} Z. Liu and E. Manousakis, Phys. Rev. B {\bf 45}, 2425 (1992).
\bibitem{Sus2} O. P. Sushkov, Solid State Communications {\bf 83}, 303 (1992).
\bibitem{Gia3} T.Giamarchi and C.Lhuillier, Phys. Rev. B {\bf 47}, 2775 (1993).

\bibitem{Kuch3} M.Yu.Kuchiev and O.P.Sushkov, Physica C {\bf 218}, 197 (1993).
\bibitem{Ede2} R. Eder, Phys. Rev. B {\bf 45}, 319 (1992).
\bibitem{Bon2} M. Boninsegni and E. Manousakis, Phys. Rev. B {\bf 47},
11897 (1993).
\bibitem{Poilblanc} D. Poilblanc, Phys. Rev. B {\bf 48}, 3368 (1993).
\bibitem{Sher} A. V. Sherman, Physica C {\bf 211}, 329 (1993).
\bibitem{PoilDag} D. Poilblanc, J. Riera, and E. Dagotto, Phys. Rev. B
{\bf 49}, 12318 (1994).
\bibitem{Cher3} A.L.Chernyshev, A.V.Dotsenko, and O.P.Sushkov,
Phys. Rev. B {\bf 49}, 6197 (1994).

\bibitem{Des3} D. S. Dessau et al., Phys. Rev. Lett. {\bf 71}, 2781 (1993).
\bibitem{Des33} D. S. Dessau, Z.-X. Shen, and D.M.Marshall, Phys. Rev. Lett.
{\bf 71}, 4278 (1993).

\bibitem{Dag4} E. Dagotto, A. Nazarenko, and A. Moreo, Phys. Rev. Lett., to be
published.
\bibitem{Suhf} O. P. Sushkov, Phys. Rev. B {\bf 49}, 1250 (1994).

\bibitem{Esk0} H. Eskes, G. A. Sawatzky, and L. F. Feiner, Physica C
{\bf 160}, 424 (1989).
\bibitem{BCh3} V. I. Belinicher and A. L. Chernyshev, Phys. Rev. B
{\bf 47}, 390 (1993).

\bibitem{Land} L. D. Landau and E. M. Lifshitz, Quantum Mechanics.
Nonrelativistic Theory (Pergamon Press, Oxford, 1965), p. 156.
\end{references}
\end{document}